\documentclass[preprint,12pt]{article}

\usepackage{arxiv}
\usepackage{graphicx,color,psfrag}

\begin{document}

\thispagestyle{empty}

\title{A Rule Search Framework for the Early Identification of Chronic Emergency Homeless Shelter Clients}

\author{  Caleb John\\
  University of Calgary  \\
  2500 University Dr.~NW, Calgary, AB, Canada, T2N 1N4\\
  ctjohn@ucalgary.ca\\
  \and
  Geoffrey Messier  \\
  University of Calgary  \\
  2500 University Dr.~NW, Calgary, AB, Canada, T2N 1N4\\
  gmessier@ucalgary.ca
}

\date{}

\maketitle

\begin{abstract}
This paper uses rule search techniques for the early identification of emergency homeless shelter clients who are at risk of becoming long term or chronic shelter users.  Using a data set from a major North American shelter containing 12 years of service interactions with over 40,000 individuals, the optimized pruning for unordered search (OPUS) algorithm is used to develop rules that are both intuitive and effective.  The rules are evaluated within a framework compatible with the real-time delivery of a housing program meant to transition high risk clients to supportive housing.  Results demonstrate that the median time to identification of clients at risk of chronic shelter use drops from 297 days to 162 days when the methods in this paper are applied.
\end{abstract}

\section{Introduction}
\label{sec:intro}

Homelessness is one of the great challenges facing society today and the approach to meeting this challenge is changing.  In Canada, a country with a population of approximately 37~million people, at least 235,000 individuals experience homelessness in a year \cite{gaetz-s-2016}.  Traditionally, emergency shelters have been a major part of supporting these individuals with 450,000 Canadians using a shelter at least once between 2010 and 2014 \cite{gaetz-s-2016}.  However, this is changing as the homelessness serving sector transitions to the Housing First model.   Many emergency shelter clients experience a variety of mental health, physical health and addictions related challenges.  Rather than expecting individuals to resolve these challenges while living in shelter, Housing First programs quickly connect people experiencing homelessness to permanent housing without pre-conditions \cite{goering-p-2014, usich-2015}.  People living with long term or {\em chronic} homelessness are a high priority group for referral to housing programs \cite{gaetz-s-2014, aubry-t-2015, toros-h-2018}.  

Identifying members of the chronic homeless population in a timely fashion is a challenge.  There are a variety of chronic homelessness definitions that can be used to identify which shelter clients fall into this group  \cite{byrne-t-2015, di-community-report-2019, synder-sg-2008, employment-social-development-2019}.  However, these definitions require a minimum of 6-12 months of homelessness before an individual is classified as chronic.  This delay places individuals at risk since conditions in most emergency shelters are difficult and have the potential to adversely affect the mental and physical health of an already very vulnerable group.

A second complicating factor is that only a minority of shelter users will become chronic.  Typically, a large majority (over 80\%) of first time emergency shelter clients will exit shelter after a short time with very little help \cite{hastie-t-2017, kuhn-r-1998, aubry-t-2013, kneebone-r-2015}.  Since supportive housing resources in most municipalities are limited, it would be inefficient to allocate these supports to a group of individuals who would exit shelter successfully on their own.

The primary contribution of this paper is to present a machine learning framework for the early identification of individuals at risk of becoming chronic shelter clients.  Our approach will identify the minority of shelter users who will become chronic as soon as possible so they can be referred to supportive housing programs.  This goal will be achieved using methods with the additional advantage of being intuitive for non-technical shelter staff and compatible with the rudimentary information technology (IT) infrastructure commonly found in most not-for-profit shelter settings.

Machine learning has been applied to the problem of homelessness in a variety of ways.  A commonly studied objective is the identification of individuals likely to enter or re-enter homelessness after exiting shelter \cite{diguiseppi-gt-2020, chan-h-2017, kube-a-2019, greer-al-2016, hong-b-2018, shinn-m-2013, gao-y-2017, byrne-t-2019}.  A smaller body of work addresses the early identification of people who will experience chronic homelessness \cite{vanberlo-b-2020, toros-h-2018}.   Logistic regression and decision trees are the most popular algorithms utilized by these authors with one study \cite{vanberlo-b-2020} exploring the use of neural networks.

Most researchers recognize that their work is more likely to be adopted if it is interpretable \cite{chan-h-2017, kube-a-2019, vanberlo-b-2020}.  Any application of machine learning in shelter will be a collaboration between the machine learning experts who develop and deploy the machine learning system and front-line shelter staff who use the system in their interactions with clients.   Shelter staff must have an intuitive understanding of the system in order to trust its decisions.  While interpretability is sometimes cited as the reason to utilize logistic regression \cite{chan-h-2017, kube-a-2019}, we argue that even logistic regression is largely opaque to anyone without an advanced mathematical background.  In contrast, the threshold style rules generated by decision trees \cite{toros-h-2018, chan-h-2017, gao-y-2017} are much more intuitive for non-technical staff.  For example, if a decision node states that the number of violent incidents experienced by a client should be less than a threshold, a non-technical shelter staff member can both understand that rule and likely validate it with their own experience. 

Machine learning will also only make an impact in the fight to end homelessness if it is compatible with emergency shelter IT infrastructure.  Our experience with 37 different housing and homelessness agencies in Calgary, Canada reveals an IT landscape consisting primarily of database style systems that support basic queries and often have closed, proprietary user interfaces.  We feel this is representative of most cities in North America.  This places most machine learning algorithms out of reach for many agencies without a significant investment in upgrading both their technical infrastructure and staff expertise.  Even the ordered nature of the rules produced by decision trees are difficult to implement on many systems.

We will meet the dual requirements of interpretability and IT compatibility by utilizing rule set learning techniques \cite{furnkranz-j-2012} for the early identification of clients at risk of becoming chronic shelter users.  While rule set learning and decision trees produce similar threshold style rules, rule sets have the important distinction of being unordered.  This makes them much easier to implement using a database interface that supports basic thresholding and logical operators.  While work has previously been done exploring threshold testing for chronic shelter use identification \cite{me-ijh-2021, me-jths-2021}, this work lacks the rigor of a formalized rule search framework that optimizes rule selection based on a performance metric.

Our second contribution is to present a system that requires only data that is consistent with an inclusive and trauma informed philosophy to emergency shelter access.  Most previous studies utilize data features collected via intake surveys \cite{diguiseppi-gt-2020, chan-h-2017, kube-a-2019, greer-al-2016, hong-b-2018, shinn-m-2013, gao-y-2017, byrne-t-2019, vanberlo-b-2020} that include answers to questions related to employment, health, social stability, past traumatic experiences, etc.  While these data features are powerful predictors, collecting survey data can be re-traumatizing for some clients.  For this reason, some of the most vulnerable individuals using shelter services refuse to participate in surveys.  The Calgary Drop-In and Rehab Centre (the DI) is our partner for this research and is an organization that imposes very few barriers to access their services.  New clients at the DI are encouraged but not required to complete an intake survey.  To ensure all clients are included in our analysis, not just those associated with survey data, our study will utilize only administrative data from the DI containing events linked to all clients after they are accepted to shelter.  Examples of these events include accessing sleep services, electing to speak with a counselor and experiencing a violent incident.

Our final contribution is to evaluate our chronic client identification performance using an approach that is consistent with how machine learning would be incorporated into the real-time delivery of an emergency shelter housing program.  Much of the existing literature utilize data collected over very long observation windows, consisting of several months to several years \cite{diguiseppi-gt-2020, chan-h-2017, kube-a-2019, greer-al-2016, hong-b-2018, shinn-m-2013, gao-y-2017, byrne-t-2019, vanberlo-b-2020, toros-h-2018}.  We assume that housing triage meetings are held on a monthly basis and that clients can only be evaluated using the data collected between their shelter entry and the meeting date.  As a result, we evaluate our methods using a framework where clients are evaluated using only the data available at successive monthly meetings.  While this degrades performance relative to very long observation windows, it is consistent with the philosophy of identifying clients for assistance as soon as possible.  We will also demonstrate that our approach performs well, even with this constraint.

In the following, Section~\ref{sec:methods} introduces the dataset used for this study and the algorithm chosen to identify chronic shelter clients.  Section~\ref{sec:results} evaluates the performance of our approach.  Discussion and concluding remarks are provided in Sections~\ref{sec:disc} and \ref{sec:concl}, respectively.  While implementing the ideas in this paper will only be possible through a multi-disciplinary collaboration between machine learning staff and front-line emergency shelter social work staff, the language and subject matter in this paper is intended for a machine learning audience.

\section{Methods}
\label{sec:methods}

\subsection{Data}
\label{ssec:data}

This is a secondary data analysis performed on anonymized shelter records collected at the Calgary Drop-In Centre (DI) between July 1, 2007 and January 20, 2020.  The data anonymization and client privacy protocol used during this study was approved by the University of Calgary Conjoint Faculties Research Ethics Board.  The dataset consists of 5,060,302 entries for 41,935 unique DI client profiles.  Each entry records a single timestamped interaction with a client that occurs within the shelter.  This interaction could be accessing sleep services (EntrySleep), accessing counseling services (EntryConsl), a security incident resulting in a ban or {\em bar} from shelter (EntryBar) or a log note (EntryLog).  To mitigate left censoring of the data, we exclude all clients who have their first sleep date appear prior to July 1, 2008.  This removes a total of 8,589 clients.    As a result, 32,346 individuals are retained in the dataset (77.1\% of the original dataset population).

Many database entries also have comments fields where shelter staff can describe an incident or consultation.  To maintain client privacy, these comment fields are stripped by DI IT staff as part of the anonymization process.  Each comment field is replaced by a series of boolean flags indicating whether certain keywords appear in that field.  Based on consultation with DI shelter staff, the keywords are divided into the following categories:  Violence, Overdose, Police/Justice, Mental Health, Physical Health,  Emergency Medical Services (EMS), Addiction, Conflict, Bar and Supports.  If one or more keywords in a category appear in a comment field, the boolean flag for that category is set to true.  For example, a common Police word was ``CPS'' (Calgary Police Services) and a common Violence word was ``fight''.  A comment field with one or more occurrences of ``CPS'' and one or more occurrences of ``fight'' would have its Police/Justice and Violence flags set to true.  All other category flags for that comment would be false.

Since this purpose of this study is to identify clients at risk of chronic shelter use, it is important to assign a flag to each individual in the database indicating whether they did in fact become a chronic shelter user over the time interval considered for this study.  This flag is determined by applying the Canadian federal government definition of chronic homelessness to each client's entire record of shelter use.  This definition states that the chronically homeless group includes ``individuals who are currently experiencing homelessness and who meet at least 1 of the following criteria: they have a total of at least 6 months (180 days) of homelessness over the past year or they have recurrent experiences of homelessness over the past 3 years, with a cumulative duration of at least 18 months (546 days)'' \cite{employment-social-development-2019}.  A total of 3,191 of the 32,346 clients (9.9\%) satisfy this definition and are designated as chronic in the dataset.

To better understand the people being considered in this study, it is also useful to compare the shelter access characteristics of the clients that do and do not meet the Canadian definition for chronic homelessness (ie. the chronic and non-chronic groups, respectively).  In the following, a {\em stay} is a 24 hour period where sleep services are accessed at least once.  An {\em episode} of shelter use is a series of stays where the separation between consecutive stay dates is less than 30 days \cite{byrne-t-2015}.  Statistical results are presented for the total shelter stays per client, total episodes per client, shelter tenure per client (number of days between first and last stay), shelter use percentage (stays divided by tenure) and the average gap length between episodes of shelter use.  Since shelter client characteristics are often exponentially distributed, we show median, 10th percentile and 90th percentile in addition to average.  The results for the chronic and non-chronic groups are provided in Tables~\ref{tb.Chronic} and \ref{tb.NonChronic}, respectively.  The statistics shown in Tables~\ref{tb.Chronic} and \ref{tb.NonChronic} are broadly consistent with other studies of chronic and non-chronic shelter users in North America \cite{kuhn-r-1998, culhane-dp-2007, aubry-t-2013, kneebone-r-2015}.

\begin{table}[htbp]
\centering
\begin{tabular}{r|c|c|c|c}
& Average & Median & 10th Pct. & 90th Pct. \\ \hline
Total Stays & 963.6 & 671.0 & 277.0 & 2140.0\\
Total Episodes & 6.9 & 5.0 & 1.0 & 15.0\\
Tenure (days) & 2254.4 & 2237.0 & 612.0 & 3960.0\\
Usage Percentage & 48.1 & 40.9 & 15.3 & 93.4\\
Avg.~Gap Length (days) & 3.2 & 2.4 & 1.0 & 6.5\\
\end{tabular}
\caption{Chronic user shelter access characteristics.}
\label{tb.Chronic}
\end{table}

\begin{table}[htbp]
\centering
\begin{tabular}{r|c|c|c|c}
& Average & Median & 10th Pct. & 90th Pct. \\ \hline
Total Stays & 34.6 & 5.0 & 1.0 & 108.0\\
Total Episodes & 3.3 & 2.0 & 1.0 & 8.0\\
Tenure (days) & 766.1 & 162.0 & 1.0 & 2590.0\\
Usage Percentage & 46.6 & 13.7 & 0.4 & 100.0\\
Avg.~Gap Length  (days) & 164.9 & 20.0 & 1.0 & 430.3\\
\end{tabular}
\caption{Non-chronic user shelter access characteristics.}
\label{tb.NonChronic}
\end{table}

\subsection{Feature Generation}
\label{ssec:features}

We assume that shelter case workers convene {\em housing meetings} on the first of each month to determine which clients should be referred to housing programs.  Case workers will base their housing decisions on all data collected for a client.  This would include the responses to survey questions, if available.  However, in order to be as inclusive as possible and avoid potential bias towards clients who elect to do intake surveys, we will restrict ourselves to using only the shelter event data described in Section~\ref{ssec:data} since it is available for all clients.

We assume that housing recommendations are made only for clients who have recently interacted with shelter services and staff.   An {\em active client} is defined as one who has accessed shelter services at least once in the 30 days prior to the housing meeting.   {\em Attribute summary tables} are prepared that summarize the shelter administration data available for each active client as of the date of the housing meeting.  These tables are a sum of the number of data entry types described in Section~\ref{ssec:data} (ie. EntrySleep, EntryLog, etc.) and the number of data entries with particular keyword flags set (ie. number of entries with the Violence keyword set, number of entries with Addiction set, etc.).

The appropriate length of time a client should be in shelter before being assessed for housing is an important question.  If this time window is too short, there may be insufficient data to accurately predict if a client will become chronic.  On the other hand, longer time windows will improve accuracy at the cost of increased time in shelter before receiving help.  To explore this tradeoff, attribute summary tables are generated for window sizes of $w \in \{30,60,90,120\}$ days.  Section~\ref{ssec:data} describes $K$ = 14 data attributes consisting of 3 one hot entry type indicators and 11 keyword boolean flags.  The attribute table for a window size of $w$ contains sums of each client's $K$ data attributes calculated at the first monthly meeting where they have accumulated at least $w$ days in shelter.  Attribute tables are only generated for clients who meet the active client criterion on the meeting date.  Each entry in an attribute table also contains a boolean flag indicating whether the client satisfies the definition for chronic homelessness described in Section~\ref{ssec:data}.  Therefore, the dimensions of an attribute summary table for window size $w$ are equal to $N_w \times (K+1)$, where $N_w$ is the number of active clients in the table.  Each row in an attribute summary table is referred to as an {\em example}.

Table~\ref{tb.NExsFtrs} shows the number of clients contained in each attribute summary table (ie. the number of examples in each table) relative to the total study population.  As expected, the number of clients included in each attribute summary table drops for the longer window sizes since most clients remain only a short time in shelter.  Some clients may appear in attribute summary tables for more than one window size.

\begin{table}[htbp]
\centering
\begin{tabular}{c|c|c}
Win.~Size & Number of Examples, $N_w$ & Num.~Features, $L$ \\ \hline
30 &  12175/32346 (37.64\%) & 491 \\
60 &  8671/32346 (26.81\%) & 655 \\
90 &  7440/32346 (23.00\%) & 779 \\
120 & 6650/32346 (20.56\%) & 903 \\
\end{tabular}
\caption{Example and feature counts.}
\label{tb.NExsFtrs}
\end{table}

A {\em feature} is the binary result of a threshold test applied to one of the $K$ attribute columns in an attribute summary table \cite{furnkranz-j-2012}.  The two threshold comparison operators are $\{ \geq, < \}$.   Each distinct combination of a comparison operator and threshold value corresponds to a different feature.  To prevent the feature space from getting too large, the feature generation algorithm presented in \cite{gamberger-d-2002} is used where only threshold examples that discriminate at least one pair of positive and negative examples are included.  The resulting features are contained in $N_w \times (L+1)$ {\em coverage tables} where $L > K$ is the number of features and an additional column containing the chronic shelter use label is retained.  The rows of each coverage table are determined by applying the $L$ feature tests to the corresponding row in an attribute summary table.  The number of features in each coverage table are presented in Table~\ref{tb.NExsFtrs}.  The number of features grows with window size since the attribute summary tables contain a larger range of values for longer windows.

A rule set consists of the conjunctive and disjunctive combination of a small subset of the $L$ features contained in a coverage table that yield the best performance according to a chosen metric like accuracy, precision, etc.  Searching for this subset among the very large number of features indicated in Table~\ref{tb.NExsFtrs} has the potential to require considerable computation.  The search space can be reduced by discarding attributes that are either highly correlated with another attribute and/or poorly correlated with the chronic shelter use outcome label.  

This attribute pruning is implemented by first determining the correlation coefficients between each of the $K$ attributes in each attribute summary table and the correlation coefficient between those attributes and the chronic homelessness outcome label.  Attributes are placed in pairs in decreasing order of cross correlation between the pair members.  For example, the two attributes with the highest cross-correlation are placed in the first pair, the remaining two with the next highest cross correlation are placed in the next pair and so on.  Then, the member of each pair that is the least correlated with the output label is discarded.  This process is repeated until all attributes have been discarded.  The reverse of the discard order defines the order that attributes should be retained.

The feature retention order results are provided in Table~\ref{tb.AttRetention} where features at the top of the table are more valuable than the features below.  The table shows that the set of most valuable features converge for window sizes greater than 30.   A particularly important note is that the top three best attributes are entry types (EntrySleep, EntryBar and EntryConsl). Database entry types are easy to access in most IT systems but keyword counts may be beyond the technical capabilities of many shelters.  A rule set that operates only by counting the number of database category entry types will be much easier to implement.

Based on Table~\ref{tb.AttRetention}, we define two attribute sets.  The {\em core} attribute set is \{EntrySleep, EntryBar, EntryConsl\}.  The {\em extended} attribute set is \{EntrySleep, EntryBar, EntryConsl, Violence, Overdose, Addiction\}.  These sets are used to generate {\em core coverage tables} and {\em extended coverage tables}, respectively.

\begin{table}[htbp]
\centering
\begin{tabular}{c|c|c|c}
$w$ = 30 days & $w$ = 60 days & $w$ = 90 days & $w$ = 120 days \\ \hline
EntrySleep      & EntrySleep  & EntrySleep      & EntrySleep       \\
Violence        & EntryBar    & EntryBar        & EntryBar        \\
Overdose        & EntryConsl  & EntryConsl        & EntryConsl      \\
EntryConsl      & Violence    & Violence      & Violence        \\
EntryBar      & Overdose      & Overdose      & Overdose         \\
Addiction      & Addiction    & MentalHealth      & MentalHealth   \\
PhysicalHealth  & PhysicalHealth & PhysicalHealth  & PhysicalHealth  \\
PoliceJustice   & PoliceJustice  & PoliceJustice   & PoliceJustice       \\
EMS             & EMS            & EMS             & EMS              \\
MentalHealth   & MentalHealth    & Addiction   & Addiction      \\
\end{tabular}
\caption{Attribute retention order.}
\label{tb.AttRetention}
\end{table}

\subsection{Rule Search Algorithm}
\label{ssec:algs}

Rule search operates by creating a disjunctive set of rules that combine to identify or {\em cover} as many positive examples in the data set as possible.  Each rule in these sets is a conjunctive combination of certain data features extracted from the coverage tables.  The ``search'' part of rule search involves finding the best possible conjunctive rule based on a metric like accuracy or precision.  These search algorithms are generally divided into two categories: heuristic and exhaustive search.  Heuristic searches are efficient at the potential cost of settling on a locally optimum rule.  Exhaustive searches are guaranteed to find the global optimum with the potential penalty of being inefficient.

The OPUS algorithm \cite{webb-g-1995} is chosen for this study since it uses exhaustive search to reliably find the globally optimum rule while improving efficiency by pruning regions of search space that represent either redundant rules or rules that will never exceed the current best rule candidate.  OPUS will optimize according to a variety of quality criteria but F-score was chosen for this study \cite{furnkranz-j-2012}.  F-score finds a compromise between precision and recall that can be adjusted according to a factor $\beta^2$ where $\beta^2$ = 0 results in optimizing solely on precision and $\beta^2 = \infty$ will optimize solely on recall.  Recall is true positive rate defined as $|\hat{P}|/|P|$, where $|\hat{P}|$ is the number of true positives identified by the rule and $|P|$ is the total number of positives.  Precision is $|\hat{P}|/(|\hat{P}|+|\hat{N}|)$, where $|\hat{P}|+|\hat{N}|$ is the total number of examples identified by the rule (true positives plus false positives).   In the context of referring shelter clients to housing programs, precision is important to be confident that scarce housing resources are being efficiently allocated to shelter users that will become chronic.  On the other hand, recall is also important to ensure that no one with the potential to become a chronic shelter user is missed.  In addition to the standard optimal search OPUS termination criterion, the algorithm is modified to terminate after searching all rules of maximum length, $M_R$.

The OPUS algorithm is wrapped in a simple disjunctive rule set learning algorithm.  OPUS is used to find a single conjunctive rule.  Then, all examples covered by that rule are removed and OPUS is re-run on the remaining examples to find a second rule that is combined disjunctively with the first.  This continues until all positive examples are covered or the rule set reaches a maximum length, $M_S$.

\subsection{Real Time Program Delivery}
\label{ssec:deliver}

Our objective is to integrate the rule set search techniques in this paper into the real time delivery of housing programs within an emergency shelter.   This process begins by having machine learning experts run the algorithms in Section~\ref{ssec:algs} on a historical shelter client dataset.  Ideally, each shelter using this system would retain a machine learning consulting firm to perform this on data from their own clients.  However, this is not strictly necessary since the rules presented in Section~\ref{sec:results} could be used as a reasonable starting point for most North American shelters.

Once the rules are determined, they are provided to shelter staff to be used in preparing a list of clients for discussion during monthly housing referral meetings.  Due to the unordered, threshold style nature of the rules, the shelter staff can implement the rules using any database or spreadsheet system that allows basic query functionality.  For each meeting, rules for a window size of $w$ days are run on the most recent $w$ days of data for all active clients.  Any clients detected by the rules are placed on a triage list for discussion during the meeting.  It is important to note that clients have multiple chances of being detected. For example, a client who perhaps fails the active client criterion or does not meet the rule set thresholds for one meeting will remain in the shelter system and will be re-evaluated for triage lists at subsequent meetings.

\section{Results}
\label{sec:results}

\subsection{Rule Search Parameter Selection}
\label{ssec:parms}

The rule search algorithm described in Section~\ref{ssec:algs} has a number of parameters that need to be selected.  This exploration begins with an initial maximum rule length and rule set size of $M_R$ = 3 and $M_S$ = 2, respectively.  While longer rules and rule sets are certainly computationally feasible, one of our main objectives is to generate rule sets that are both intuitive for shelter workers to understand and practical to enter by hand into database query tools.  Selecting modest values for $M_R$ and $M_S$ achieves this objective.  The following explores the rule search parameter settings that find a compromise between performance, interpretability and search efficiency.  The results in this section are determined using stratified K-fold cross validation \cite{hastie-t-2017} with 10 folds.

\paragraph*{F-Score $\beta^2$ Setpoint}  Values of $\beta^2 \in \{ 0.01, 0.25, 0.5, 1 \}$ are evaluated with the extended attribute set described in Section~\ref{ssec:features} and a $w$=90 day window size.  The $\beta^2$ values yield (precision, recall) values of (0.8644, 0.1978), (0.6254, 0.6793), (0.5236, 0.7978) and (0.5273, 0.8569), respectively.  As expected, increasing $\beta^2$ improves recall at the cost of precision with the value of $\beta^2$ = 0.25 yielding a good compromise between the two parameters.  

\paragraph*{Extended vs. Core Attribute Set}  Values $\beta^2$ = 0.25 and a $w$ = 90 day window size are then used to explore how simplifying the number of input attributes affects performance.  The core and extended rule sets are compared for $M_R$ = 3 and $M_S$ = 2 which yields (precision, recall) values of (0.6260, 0.6832) and (0.6254, 0.6793), respectively.  Since these results are very comparable, the core attribute set is used for the remainder of this paper.  This has the advantage of reducing search times and, as noted in Section~\ref{ssec:features}, uses features easily extracted from most shelter databases.

\paragraph*{Effect of Rule Length} Next, reducing rule set size and rule length is explored for the core attribute set, $\beta^2$=0.25 and a 90 day window size.  Values of $M_R$=3 and $M_S$=1 result in (precision, recall) values of (0.765, 0.5034).  As expected, $M_S$=1 reduces recall since the disjunctive combination of multiple rules works to include more positive individuals.  Values of $M_R$=2 and $M_S$=1 result in comparable (precision, recall) values of (0.7427, 0.5336).  The values of $M_R$=2 and $M_S$=1 are used for the remainder of this paper since shorter rules have the practical benefit of both reducing search time and being very simple for shelter staff to work with.  As we will demonstrate in the next section, using the rules in the monthly triage framework described in Section~\ref{ssec:deliver} will recover the loss in recall that results from choosing $M_S$=1.

\paragraph*{Effect of Window Size}  Finally, the overall performance for the different window sizes can be evaluated using $M_R$ = 2, $M_S$ = 1, $\beta^2$ = 0.25 and the core attribute set.  Table~\ref{tb.WinPrf} shows the performance estimated using cross-validation along with the rules that result from training on the entire coverage table for each window size.  Rule performance does improve with longer window size but not dramatically so.  A window size of 90 days appears to offer the best compromise between rapid identification and performance.

\begin{table*}[htbp]
\centering
\begin{tabular}{c|c|c|c}
Window Size (days) & Precision & Recall & Rule\\ \hline
30 &  0.4611 &  0.4514 & EntrySleep $\geq$ 28 $\cap$ EntryBar $<$ 0.5 \\
60 &  0.6092 & 0.4744 & EntrySleep $\geq$ 54 $\cap$ EntryConsl $<$ 10.5\\
90 & 0.7427 &  0.5336 & EntrySleep $\geq$ 78 $\cap$ EntryBar $<$ 3.5\\
120 & 0.7979 &  0.5362 & EntrySleep $\geq$ 99 $\cap$ EntryBar $<$ 4.5\\
\end{tabular}
\caption{Window size performance.}
\label{tb.WinPrf}
\end{table*}

\subsection{Real Time Program Delivery}
\label{ssec:delver_results}

This section evaluates the performance of the rules in Table~\ref{tb.WinPrf} when integrated into a monthly housing program triage process as described in Section~\ref{ssec:deliver}.  Unlike the results in Section~\ref{ssec:parms} which are generated using coverage tables containing only the clients indicated in Table~\ref{tb.NExsFtrs}, the results in this section are generated using the entire population in the DI shelter data set.  Table~\ref{tb.PrgPrf} shows the resulting precision, recall and confusion matrix values.  The table also shows median time to identification (TTI) for all chronic clients.   TTI is the number of days a chronic client spends in shelter before being identified by the rules in Table~\ref{tb.WinPrf} for referral to a monthly housing meeting.  TTI for those clients who are never detected by a rule is calculated using the first day they satisfy the Canadian definition of chronic homelessness described in Section~\ref{ssec:data}.

\begin{table}[htbp]
\centering
\begin{tabular}{c|c|c|c|c|c|c|c}
Win  &        &           & True & False & False & True & Median \\
Size & Recall & Precision & Pos. & Neg.  & Pos.  & Neg & TTI (days) \\
\hline
30 & 0.85 & 0.60 & 2726 & 465 & 1792 & 27363 & 162.00 \\
60 & 0.77 & 0.71 & 2452 & 739 & 991 & 28164 & 225.00 \\
90 & 0.75 & 0.78 & 2409 & 782 & 680 & 28475 & 254.00 \\
120 & 0.73 & 0.83 & 2338 & 853 & 465 & 28690 & 272.00 \\
\end{tabular}
\caption{Program delivery performance.}
\label{tb.PrgPrf}
\end{table}

\section{Discussion}
\label{sec:disc}

Table~\ref{tb.PrgPrf} demonstrates that rule search methods for identifying individuals at risk of chronic shelter use provide very strong performance when integrated into a monthly housing triage program.   The authors in \cite{vanberlo-b-2020} achieve similar results using neural networks but rule sets are much easier to adopt in shelter, as noted in Section~\ref{sec:intro}.  The results in Table~\ref{tb.PrgPrf} are also much better than Table~\ref{tb.WinPrf}.  Table~\ref{tb.WinPrf} is produced using the subset of the client population used to generate the coverage tables.  This subset consists only of those clients who meet the active client criterion after their first $w$ days of being in the client database.   This is a more challenging dataset since this sub-population contains a higher proportion of clients with heavy initial shelter use but who still do not become chronic.  
Table~\ref{tb.PrgPrf} is produced using the entire population in the data set which contains more easy to distinguish individuals and results in a performance improvement.

The best choice of window size is $w$ = 30 days which achieves the best recall and median TTI.  Ideally, TTI would be even lower but 162 days is still a considerable improvement over using the Canadian federal definition of chronic homelessness to identify clients which would result in a median TTI of 297 days for this data set.  The high recall of using $w$ = 30 days ensures that as many vulnerable clients as possible are referred to housing.  However, this comes at the cost of also having the lowest precision in Table~\ref{tb.PrgPrf}.

A low precision raises two potential concerns: the false positives may overwhelm housing program capacity and/or the false positives are not good candidates for limited housing resources.  Regarding program capacity, the 30 day rule set identifies a total of 4,518 clients in a data set that stretches over 101 months.  This corresponds to 44.7 clients per month which the DI has already demonstrated is within the capacity of their housing programs \cite{di-community-report-2019}.   Regarding whether the false positives are good candidates for housing, Table~\ref{tb.FPos} shows the shelter access statistics for the 1,792 false positive clients identified by the 30 day rule.  While these individuals are in shelter less time than the chronic clients in Table~\ref{tb.Chronic}, they are still very heavy shelter users.  With median shelter stays of over 150 days and tenures of interaction over 1000 days, we feel that this false positive group are still excellent candidates for housing.  Referring these individuals for support would not be a misallocation of resources.

\begin{table}[htbp]
\centering
\begin{tabular}{r|c|c|c|c}
& Average & Median & 10th Pct. & 90th Pct. \\ \hline
Total Stays & 186.3 & 153.0 & 72.0 & 350.0\\
Total Episodes & 6.1 & 4.0 & 1.0 & 14.0\\
Tenure (days) & 1493.0 & 1210.0 & 122.0 & 3410.0\\
Usage Percentage & 30.0 & 16.1 & 5.0 & 86.0\\
Avg.~Gap Length  (days) & 8.9 & 6.2 & 1.1 & 20.1\\
\end{tabular}
\caption{False positive shelter access characteristics.}
\label{tb.FPos}
\end{table}

Interestingly, all of the rules in Table~\ref{tb.WinPrf} have the same structure: they are looking for clients who have a lot of sleep stays in shelter and very few bar or counseling events.  It is not surprising that accumulating a lot of shelter stays is a strong indicator of future chronic shelter use.  However, imposing an upper limit on bars and counseling is more interesting.  Anecdotally, DI shelter staff indicate that chronic shelter users often ``fly under the radar''.  They accumulate a lot of shelter stays but otherwise avoid attracting attention.  A bar event tends to occur for clients who are barred from accessing shelter for a period of time due to being violent or otherwise breaking shelter rules.  Counseling services are associated with clients actively trying to exit shelter.  Clients that maintain a low profile would naturally not accumulate many of either type of event.  Low profile clients are also exactly the clients that benefit from a machine learning identification system, since they will not necessarily be top of mind due to their limited interactions with staff and may otherwise ``fall through the cracks''. 

It is important to note the limitations of performing this type of study on data from a single emergency shelter.  First, our results are necessarily biased towards individuals who make use of shelter services.  Many people living with homelessness sleep outside and would not be captured in our data.  This shortcoming can be addressed by applying the techniques in this paper to a data set that has been augmented to include individuals sleeping outside, possibly using input from street outreach teams.  Cluster analysis has previously demonstrated that the client shelter access patterns in the DI data set are broadly representative of most North American emergency shelters \cite{me-ijh-2021}.  While the rules presented in this paper should be a reasonable starting place for most shelters, it is still best for each shelter seeking to apply our techniques to first conduct a new rule search on their own client data.  Finally, rather than keeping the rule set static, we recommend that the rules are re-examined and potentially refreshed on a regular basis.  This could be done by having the rule search algorithm re-run on an updated data set.  The thresholds could also be adjusted by the shelter staff based on their experience with the suitability of the clients on past lists.

\section{Conclusions}
\label{sec:concl}

Homelessness is one of the most significant challenges facing society today.  The Housing First philosophy recognizes that people experiencing homelessness are often struggling with many challenges and that placing them in a stable housing situation is the first step towards addressing them.  The data science community has an obligation to do its part to assist with this effort.

This paper has presented a rule search framework for the early identification of emergency shelter clients who have the potential to become long term shelter users.  Our approach achieves performance similar to neural networks and has a median time to identification of 162 days compared to 297 days when standard chronic homelessness definitions are used to identify high priority housing candidates.  We are the first to discuss in detail how machine learning can be integrated into real-time housing program delivery.  Finally, our approach is far simpler than any of the machine learning work in the homelessness sector to date, will make intuitive sense to non-technical staff and will be compatible with the basic IT infrastructure found in most not-for-profit shelters.  This increases the likelihood of our approach making the transition from the research domain to practice.

Finally, we conclude by expressing our strong opinion that it would be a mistake to rely exclusively on any machine learning method for supportive housing referrals.  Chronic shelter users are extremely complex and benefit from a range of assessment and triaging methods \cite{aubry-t-2015}.  We feel the techniques presented in this paper are best used to identify low profile clients who may otherwise escape notice by staff, particularly in very busy shelters.  However, the final decision regarding which individuals are selected for housing and where they are placed should be made by experienced case managers working collaboratively with the clients themselves.

\section{Acknowledgments}
\label{sec:ack}

The authors would like to acknowledge the support of the Natural Sciences and Engineering Research Council of Canada (NSERC), the Calgary Drop-In Centre and the Government of Alberta.  This study is based in part on data provided by Alberta Community and Social Services.  The interpretation and conclusions contained herein are those of the researchers and do not necessarily represent the views of the Government of Alberta.  Neither the Government of Alberta nor Alberta Community and Social Services express any opinion related to this study.


\bibliographystyle{elsarticle-num}
\bibliography{plwh,Books,data,Me}


\end{document}